# Phononic Skyrmions

Liyun Cao*, Sheng Wan, Yi Zeng, Yifan Zhu, Badreddine Assouar*

Université de Lorraine, CNRS, Institut Jean Lamour; Nancy, 54000, France.

*Corresponding author emails: liyun.cao@univ-lorraine.fr, badreddine.assouar@univ-lorraine.fr

## Abstract

Skyrmions with topologically stable configurations have shown a promising route toward magnetic and photonic materials for information processing due to their defect-immune and low-driven energy. However, the practical application of magnetic skyrmions is severely hindered by their harsh cryogenic environment and complex carriers. In addition, the narrowband nature of magnetic and photonic skyrmions leads to lower data rate transmissions, restricting the development of high-speed information processing technologies. Here, we introduce and demonstrate the concept of phononic skyrmion as new topological structures to break the above barriers. The phononic skyrmion can be produced in any solid structure at room temperature, including chip-scale structures, with high robustness and ultra-bandwidth, which could pave a new path for high-speed and topological information processing technologies. We experimentally demonstrate the existence of phononic skyrmion formed by breaking the rotational symmetry of the three-dimensional hybrid spin of elastic waves. The frequency-independent spin configuration leads to the remarkable ultra-broadband and tunable feature of phononic skyrmions. We further experimentally show the excellent robustness of the flexibly movable phononic skyrmion lattices against local defects of disorder, sharp corners, and even rectangular holes. Our research also opens a vibrant horizon towards an unprecedented way for elastic wave manipulation and structuration by spin configuration, and offers a promising lever for alternative phononic technologies, including quantum information, biomedical testing, and wave engineering.

**Main**

Skyrmion, a topologically stable three-component vector field, was initially developed in elementary particles and has since been demonstrated in condensed-matter systems (1) and helimagnetic materials (2-4). The skyrmions, characterized by a real-space nontrivial topological number, were considered as a promising route toward high-density magnetic materials for information storage and transfer (5-8), due to their defect immune and low driven energy (9, 10). Such topological skyrmions have been recently extended to photonics based on dynamic electromagnetic fields with axial evanescent waves (11, 12) and spin-orbit coupling in the evanescent fields (13, 14). This shows a promising horizon (11-14) for robust photonic information processing, sensing, and lasing. However, the practical application of magnetic skyrmions is severely hindered by their harsh cryogenic environment and complex carriers. In addition, the narrowband nature of magnetic and photonic skyrmions lead to lower data rate transmissions, restricting the development of high-speed information processing technologies. To date, the nontrivial skyrmion configurations have remained untapped for phononic materials, classically known as periodic solids supporting elastic waves (15-18), due to their more sophisticated polarization states.

Phononic materials are widely used in high signal-to-noise information processing (19, 20), high-sensitive remote sensing (21), and wireless communication (22). They present potential information carriers in quantum applications (23-28), due to their unique advantages, including orders of magnitude lower phononic wavelength in comparison with photonic systems (16, 19), scalability toward integrated devices (29), anti-jamming capability (30) (negligible crosstalk between devices and with the environment), and extreme low losses (19). Thus, in actual solid carriers with ubiquitous defects, realizing a new topological robust mode, i.e., phononic skyrmion, could lead to transformative phononic applications, especially in a generally concise configuration that can be scaled accordingly for future chip-scale technologies.

Elastic waves describe the basic dynamic principle of the deformation (i.e., the three-dimensional displacement field $\mathbf{u}$) of solids in a periodic form (31, 32), reflecting the properties from classical solid motion to lattice oscillation in a quantum field. Helmholtz theorem unveiled that $\mathbf{u}$ can be decomposed into curl-free longitudinal waves $u_L$ and divergence-free transverse waves $\mathbf{u}_T$. A large variety of researches has shown that transverse waves (i.e., optical waves) possess spins characteristic (33-35), which support the nontrivial Berry phase and quantum spin Hall effect (36). Recently, the hybrid spin induced by mixed transverse–longitudinal waves, which is responsible for abnormal phenomena beyond pure transverse waves, has been uncovered in the elastic phononic system (37). The hybrid spin can inspire strong spin-momentum locking of the elastic edge modes. However, the trivial topological invariant of the latter is not robust against defects (38). Here, we construct a new nontrivial topological structure of ultra-broadband phononic skyrmion based on the three-dimensional hybrid spin of elastic waves.

## Formation of phononic skyrmions by intrinsic hybrid spins

The elastic spin arises from the hybridization between transverse and longitudinal waves in an elastic interface. The interface can be a free surface of a plate-like structure supporting the classical Lamb waves, or a free surface of a semi-infinite (for wavelength) bulk structure supporting the Rayleigh surface waves. Note that the hybrid spin does not exist in an infinite isotropic bulk space without interfaces because decoupled transverse and longitudinal waves are two independent two-dimensional wave fields.

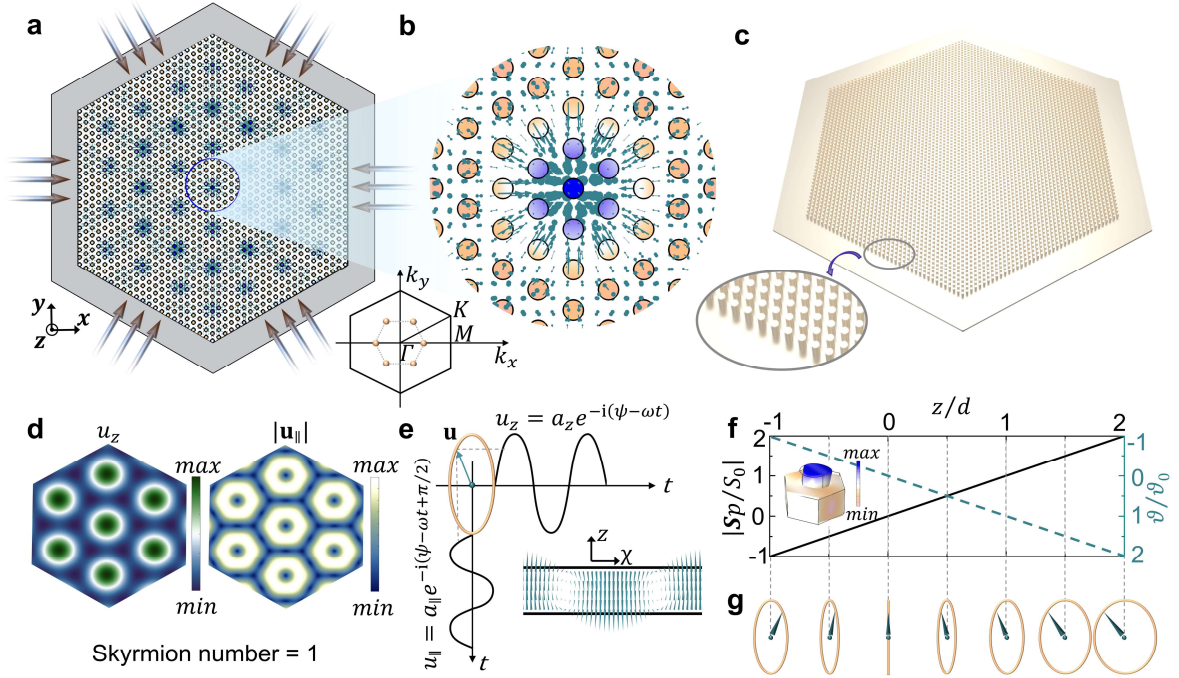

Fig. 1. **Formation of phononic skyrmions. a**, The interference of three pairs of counterpropagating plane Lamb waves breaks the rotational symmetry of intrinsic spin to construct the phononic skyrmion lattice in a hexagonal elastic meta-plate. **b**, Enlarged view of the local vector field. **c**, The meta-plate with periodically pillared resonators. **d**, The axial (out-of-plane) $u_z$ and transverse $|\mathbf{u}_\parallel|$ (in-plane) fields of the skyrmion lattices. **e**, The inset is a spin vector field at the plane formed by the $z$-axis and $\chi$-axis (same with the direction vectors $\mathbf{k}$). The axial $u_z$ and transverse $u_\parallel$ fields have a phase difference of $\pi/2$, creating to a spiral spin geometry with an elliptical trajectory. **f**, The inset shows a unit structure of an oscillating non-resonant pillared resonator on the hosting plate with the thickness of $h$. The resonator with the height of $l$ (here $l = h/2 = d = 0.5$ mm at the frequency of 8 kHz) can linearly amplify spin angular momentum $\mathbf{S}_p$ and the polarization $\vartheta = a_\parallel/a_z$ along the $z$-axis ($-d<z<2d$). Coordinate origin is on the neutral plane of the hosting plate. The $S_0$ and $\vartheta_0$ are the spin angular momentum magnitude and the polarization

on the plate surface (i.e., $z = d$). **g**, The elliptical spin trajectories of different particles along the plate thickness.

To make the structure compact, we have used a thin plate model supporting Lamb waves to build the phononic skyrmion system. The three-dimensional hybrid spin field of plane Lamb waves can be expressed by the axial (out-of-plane) and transverse (in-plane) field components

$$u_z = a_z(z) e^{-i(k \cdot \mathbf{r} - \omega t)}$$
$$\mathbf{u}_\parallel = \frac{a_\parallel(z) \cdot \mathbf{k}}{|\mathbf{k}|} e^{-i\left(k \cdot \mathbf{r} + \frac{\pi}{2} - \omega t\right)}, \tag{1}$$

where $a_z$ and $a_\parallel$ are real amplitude components (detailed derivations can be found in Supplementary Text section 1). $\mathbf{r}$ and $\mathbf{k}$ are the position and direction vectors of the transverse field, respectively. The motion of any particle $(\mathbf{r}, z)$ in the three-dimensional plate obeys an elliptical spin trajectory of $\dfrac{\text{Re}^2(u_z)}{a_z^2} + \dfrac{\text{Re}^2(u_\parallel)}{a_\parallel^2} = 1$ (Fig. 1e), whose geometry is determined by the polarization $\vartheta = a_\parallel / a_z$ (Fig. 1g). The larger the polarization, the closer the elliptical trajectory is to a circle one. In particular, at low frequencies, the lowest-order antisymmetric Lamb mode (A0) approximates the flexural wave mode (31). The corresponding spin angular momentum

$\mathbf{S}_p = -\boldsymbol{\varepsilon} \dfrac{\rho \omega}{2} \left| \text{Im} \left( \mathbf{u}_\parallel^* u_z - u_z^* \mathbf{u}_\parallel \right) \right| \approx -z k \rho \omega a_0^2 \boldsymbol{\varepsilon}$ is linearly distributed along with the plate thickness

(along the range of $-d < z < d$ in Fig. 1f), and $\boldsymbol{\varepsilon}$ is a three-dimensional unit vector perpendicular to the plane formed by the $z$-axis and the direction vector $\mathbf{k}$. Interestingly, we find that non-resonant pillared resonators (shown in the illustration of Fig. 1f) can linearly amplify $\mathbf{S}_p$ of the hosting plate due to the continuous rotation geometry (see 3D time-resolved spin vector field in Supplementary Video 1). This opens up a new degree of freedom in tuning spin texture (more details in Supplementary Text). As shown in Fig. 1f, $\mathbf{S}_p$ of the Lamb wave is linearly amplified along the range of $d<z<d+l$, by the pillared resonator with the height of $l$.

To produce the phononic skyrmions, we have designed an hexagonal meta-plate with pillared resonators (Fig. 1c), and excited three pairs of counterpropagating plane Lamb waves along with the directions of $\theta = 0°$, $60°$, and $-60°$, as shown in Fig. 1a. The corresponding wavenumber of these excited plane waves can be presented by six points in the momentum space (the lower right corner of Fig. 1a). The interference of these wave fields breaks the rotational symmetry of their spins and constructs the three-dimensional skyrmion lattice configuration of real wavefield $\overline{\mathbf{u}} = \begin{pmatrix} \overline{u}_x & \overline{u}_y & \overline{u}_z \end{pmatrix}$, which is expressed by (detailed derivations can be found in Supplementary Text section 5)

$$\overline{\mathbf{u}} = \sum_{\theta = -\frac{\pi}{3}, 0, \frac{\pi}{3}} \left[ \frac{a_{\parallel}(z) \cdot \mathbf{k}}{|\mathbf{k}|} \sin(k\mathbf{tr}), \ a_z(z) \cdot \cos(k\mathbf{tr}) \right] \cos(\omega t). \qquad (2)$$

The corresponding axial and transverse fields in a two-dimensional $x$-$y$ plane (Fig. 1d), calculated by $\overline{u}_z$ and $\sqrt{\left( \overline{u}_x \right)^2 + \left( \overline{u}_y \right)^2}$ respectively, clearly show the skyrmion lattice with the hexagonal symmetry feature. The lattice constant $a_s = \lambda / \sin(\pi/3)$ is determined by the Lamb wave wavelength $\lambda = 2\pi / k$. The axial field of the lattice varies progressively from the central "up (down)" state to the edge "down (up)" state. The zero-amplitude point of the transverse field at the center of each lattice corresponds to the polarization singularity, one typical feature of topological defects. The nontrivial real-space topology configuration can be characterized by the skyrmion number $S$

$$S = \frac{1}{4\pi} \iint_A s \, dx \, dy \,, \qquad (3)$$

where $s = \mathbf{n} \cdot \left( \partial_x \mathbf{n} \times \partial_y \mathbf{n} \right)$ is skyrmion number density, and $\mathbf{n} = \overline{\mathbf{u}} / |\overline{\mathbf{u}}|$ is the three-dimensional unit vector field. Area $A$ covers one single lattice in the $x$-$y$ plane. For such a lattice in Fig. 1d, the

calculated skyrmion number $S = 1$ confirms the nontrivial topology, leading to the robustness of the skyrmion field.

**Ultra-broadband and tunable characteristics of phononic skyrmions**

For classical waves, constructed photonic (11) or acoustic (39) skyrmions require an axial evanescent wave field to create three-dimensional wave fields. Once their carrier structure is designed, these skyrmions have a limited bandwidth since evanescent waves in the near field are only visualized in a narrow frequency band. Here, the phononic skyrmion has an ultra-broadband topological robustness feature, due to the frequency-independent three-dimensional hybrid spin texture of the elastic wave (A0 mode).

The horizontal solid line in Fig. 2a shows that the theoretical skyrmion number of the A0 wavefield in the hosting plate is always equal 1 in the wide frequency band below the first-order cutoff frequency-thickness product $fh < fh_{c1} = c_T^2/c_L$ (see detail in Supplementary Text section 2), where $c_T$ and $c_L$ are the wave speeds of the bulk transverse and longitudinal waves, respectively. A smaller thickness $h$ determines the wider frequency band. When the $fh$ is greater than $fh_{c1}$, it will introduce the first-order antisymmetric Lamb (A1) mode (see the band-like dispersions in Fig. 2b) to destroy the symmetry of the perfect skyrmion field. We point out that the imperfect skyrmion configuration, annihilated by a small-amplitude A1 mode, still has robustness due to the topological protection.

The skyrmion number density contrast $\delta = (s_{max} + s_{min})/(s_{max} - s_{min})$ is a parameter that provides a quantitative measure of the spatial confinement of the skyrmion density (11). The magenta solid line of Fig. 2a shows that the frequency-thickness product $fh$ can tune $\delta$ of the hosting plate ($l = 0$). The physical nature of the tunable characteristic is that as the $fh$ increases, the polarization $\vartheta$ (shown in Fig. 2b) of the excited spin trajectory increases. The latter determines the skyrmion texture with a large $\delta$ due to the geometrical property. For point I with a small $fh$ in Fig. 2a, the corresponding $\vartheta$ approaches 0, i.e., the spin trajectory approximates an up-and-down oscillation ($z = 0$ in Fig. 1g). This leads to bubble-type skyrmions with clear domain walls separated between two specific field states (the vector field in Fig. 2d). Note that despite the dominance of the axial field, the skyrmion number is still 1, because the non-zero transverse field always exists due to the geometry property. For point II with a large $fh$ in Fig. 2a, the large $\vartheta$ degenerates the domain walls and converts the bubble-type skyrmions to the Néel-type-like skyrmions (the theoretical vector field in Fig. 2e). The corresponding skyrmion number density profiles are shown in Figs. 2g and 2h.

The pillared resonator acts as a new degree of freedom capable of linearly increasing the polarization of the spin trajectory while maintaining the spin configuration (Fig. 1f). Therefore, by introducing pillared resonators, the wavefields in the hosting plate can keep the skyrmion configurations while the $\delta$ increases (the magenta dotted line of Fig. 2a). This can be realized if the excitation frequency is below the resonance frequency $fh_r$ (marked in the band-like dispersions in Fig. 2c) (40) to hold the wavefield symmetry. Figures 2f and 2i show the theoretical vector field and $\delta$ profile at the top plane of pillared resonators where the actual wave field exists only at the region of the pillar position. This wavefield configuration is also robust due to the topological protection of the coupled wavefield in the hosting plate. Interestingly, these topological pillared resonators can couple well with the liquid through their bending vibrations. The fluid-solid coupling leads to a nontrivial acoustic skyrmion field in liquids, which has the same ultra-broadband and tunable characteristics as the phononic skyrmion field (see details in Fig. S8). This new mechanism of generating the acoustic skyrmion break through the narrow-band property of the acoustic skyrmion constructed by evanescent waves (39).

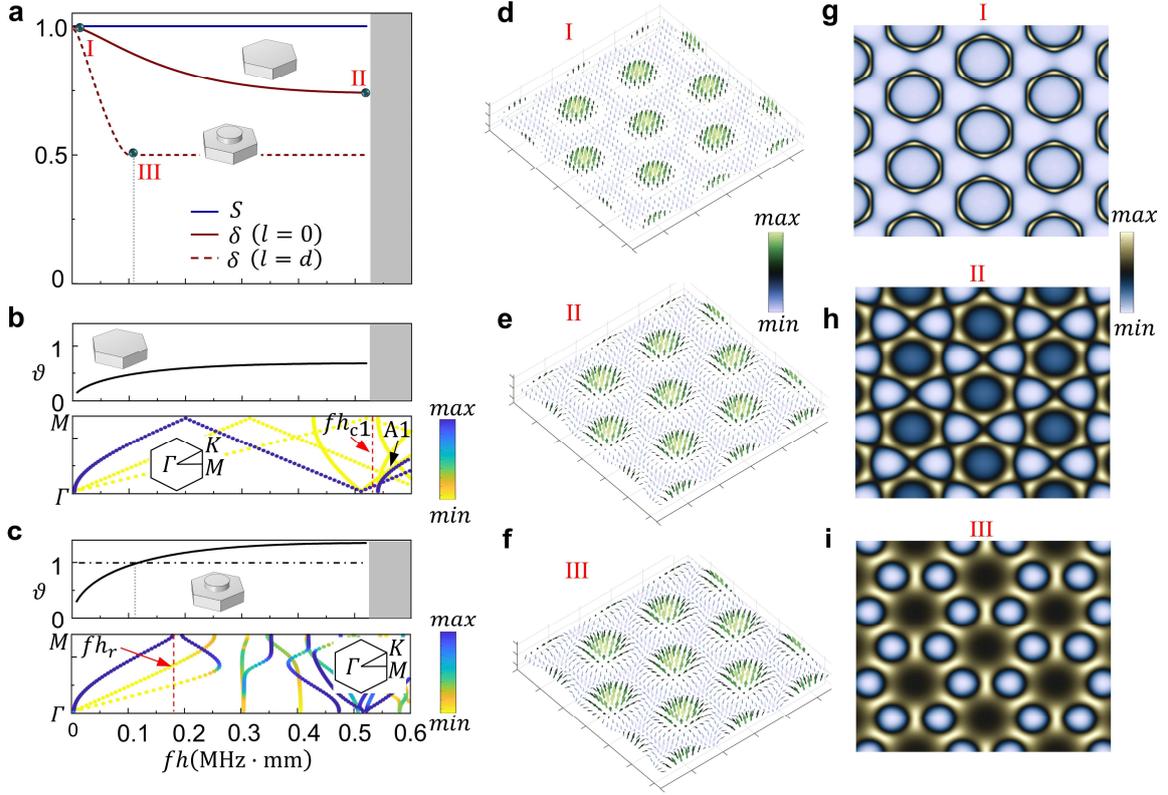

Fig. 2. **Ultra-broadband and tunable characteristics of phononic skyrmions. a**, The skyrmion number (solid blue line) of an elastic plate as function of the frequency-thickness product $fh$. The solid and dotted magenta lines indicate the skyrmion number density contrast of the wavefields in the top plane of elastic structure without and with pillared resonators, respectively. The illustrations are two single cell models, a plate with thickness of $h$=1 mm and a pillar resonator with height of $l = d$ on the plate top. **b** and **c**, The top subfigures are the polarization $\vartheta$ of the spin trajectory of the excited Lamb wave, and the bottom ones are band-like dispersions in the $\Gamma M$ direction. The pillar height is $l = d$, and the lattice constant of the single cell model is $a_0 = 4h$. **d-f**, Vector fields (color bar for the amplitude of its axial field). **g-i**, Theoretical skyrmion number density profile obtained from the theoretical calculations.

**Observation of elastic phononic skyrmions**

We have used a 3D printer to fabricate an hexagonal-symmetry meta-plate with pillared resonators, as shown in Fig. 3a. Single-side piezoelectric patch arrays excite three pairs of counterpropagating plane Lamb waves (standing wavefields) along with the directions of $\theta = 0^\circ$, $60^\circ$, and $-60^\circ$. We have used Polytec scanning laser vibrometer (PSV-500) to measure the three-dimensional dynamic displacement field on the back side of the meta-plate (41). Without loss of generality, we have measured the three-dimensional wavefield in the frequency range from 6 kHz to 9 kHz. Theoretically, perfect skyrmion wavefields can be measured at any frequency below the first-order cutoff frequency of high-order Lamb waves, even for ultra-low frequencies when the structure size is large enough. Before measuring the skyrmion wave field, we first only turn on one piezoelectric patch array (i.e., $1^{\#}$ array in Fig. 3a) to excite a plane A0 wave. The corresponding band-like experimental dispersions, shown in Fig. 3b, are obtained by Fourier transform of the scanned axial displacement field. One can observe a good agreement between the experimental and theoretical dispersions (colors and lines, respectively, in Fig. 3b). The dominance of the A0 mode indicates that the symmetry Lamb modes (S0) from the mode conversion of the non-resonance pillared resonators can be neglected, which ensures the wavefield symmetry.

For all the measured frequencies, we have calculated the skyrmion numbers by the measured wavefields of the central lattices based on Eq. (3). These experimental results, shown in Fig. 3c, have a good agreement with the theoretical ones, which confirms the broadband characteristic of the phononic skyrmions. Taking an example of skyrmion configuration in 8 kHz, we have shown the corresponding axial (out-of-plane) and transverse (in-plane) displacement fields in Figs. 3d and 3e, respectively. One can see a good agreement between the theoretical and experimental results, especially for the central skyrmion lattice. Indeed, the slight distortion for the six other skyrmion lattices around the central one is mainly due to the fact that the excited wavefields are imperfect plane waves in our compact structure. Figure 3f illustrates the two-dimensional FFT transform of the experimental axial displacement field in Fig. 3d, which shows a clear hexagonal symmetry map.

We have reconstructed 3D time-resolved Néel-type skyrmion vector field (see Supplementary Video 3). We have extracted the skyrmion number as function of the time delay in Fig. 3g for two cycles of the skyrmion wavefield. The skyrmion wavefield emanates from breaking the rotational symmetry of intrinsic spins by wavefield interference. The wavefield interference forms standing wavefields. Because the up-and-down oscillation of the standing wavefield reverses the vector direction of the three-dimensional wavefield, measured $S$ are 1 and −1 in the first and second half of the wave cycle, respectively, which is consistent with the theoretical ones (calculated from an analytical model shown in Supplementary Text section 5). Note that the measured $S$ are far from 1 and −1 in the middle of a cycle because, at these moments, the skyrmion wavefield displacement approaches 0 due to oscillation of the standing wave, which amplifies the error brought by the imperfect symmetry of attached excitation source and measurement noise. For the moment (delay time $\tau$ = 22.064 ms) of maximum axial displacement of the skyrmion wavefield, the snapshot of Supplementary Video 3 exhibits a distinct Néel-type skyrmion configuration in the three-dimensional space, as shown in Fig. 3h. The vector fields at the two-dimensional bottom plane of the hosting plate (bubble-type skyrmion) and the top plane of pillared resonators (Néel-type-like skyrmion) are shown in Figs. 3j and 3i, respectively. The transition of the different skyrmion types in the geometric space demonstrates that the pillared resonators can contribute to construct three-dimensional skyrmion configurations.

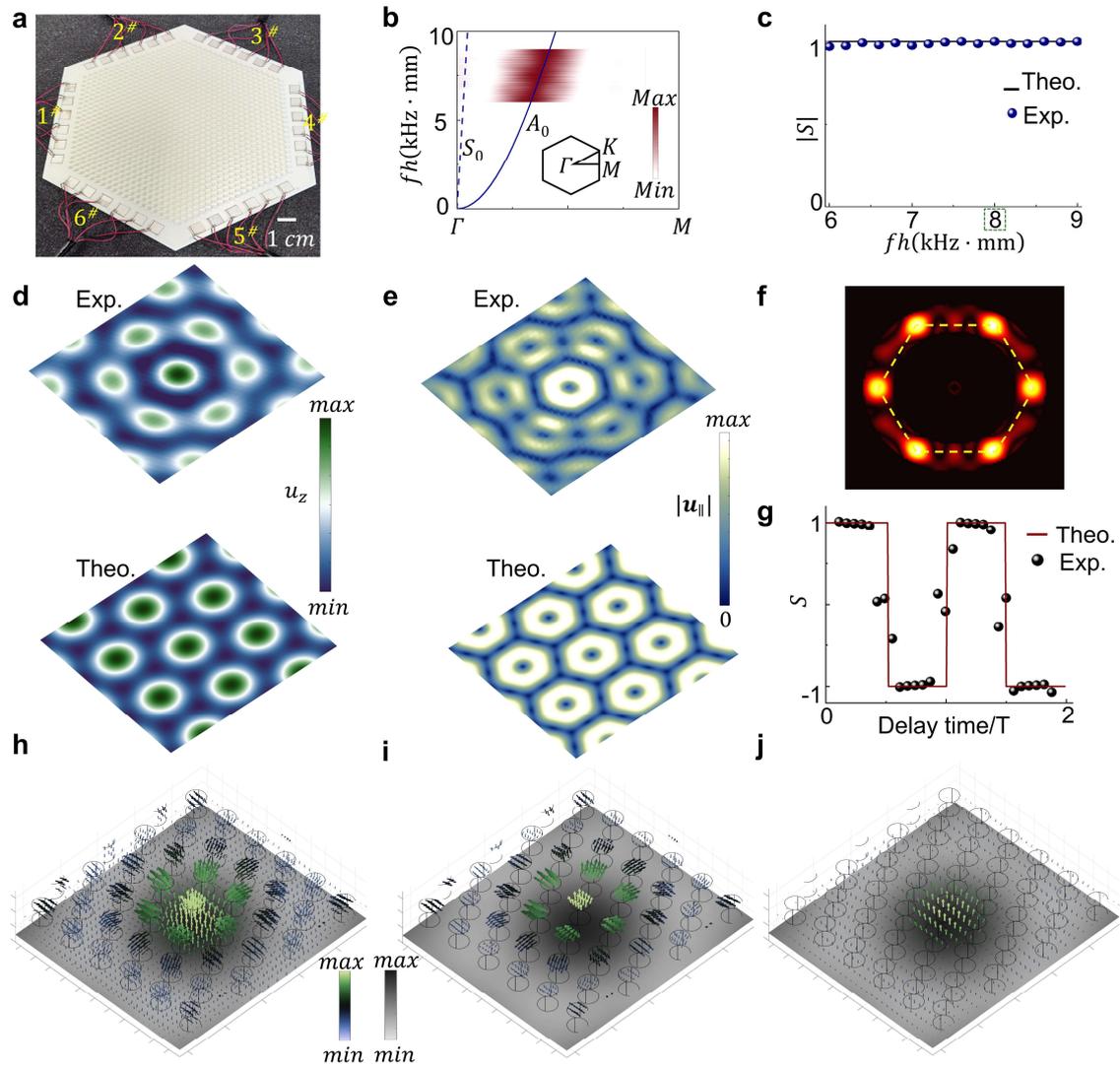

Fig. 3. **Observation of phononic skyrmions. a**, The experimental meta-plate. **b**, The experimental and theoretical band-like dispersions (colors and lines, respectively) are obtained by Fourier transform of the axial displacement field. **c**, The experimental and theoretical skyrmion number in the whole measure frequency range. **d** and **e**, The axial and transverse displacement fields in 8 kHz. **f**, The two-dimensional FFT transform of the axial displacement field. **g**, The experimental and theoretical skyrmion number as function of time delay for two cycles of the skyrmion lattice in 8 kHz. T = 0.125 ms is one cycle. **h**, The snapshot of Supplementary Video 3 for the experimental Néel-type skyrmion vector field in the three-dimensional space. The image beneath

the vectors shows the vertical component of the wavefield. **i**, The vector fields at the two-dimensional top plane of pillared resonators (Néel-type-like skyrmion). **j**, The vector fields at the bottom plane of the hosting plate (bubble-type skyrmion).

## Robustness of phononic skyrmions

The nontrivial topological phononic skyrmion lattice leads to the robustness of wavefield texture against defects. As shown in Fig. 4a, we have introduced defects into the reference structure (illustrated in Fig. 3a) by removing and disorderly arranging the pillared resonators near the structure center, marked as structure I. The defect location can be in any region of the structure, including its center (see details in Fig. S10). These defects lead to wave scattering and thus distorting the wavefield. However, the measured wavefield in the structure I (Fig. 4d) clearly demonstrates that the skyrmion lattice is hardly affected by these defects, compared with that of the reference structure (i.e., the one in Fig. 4f). To go further, we have introduced a strong defect into the structure by drilling a rectangular hole with sharp corners, as shown in Fig. 4b, marked as structure II. We have experimentally observed that the skyrmion lattices in this structure II (Fig. 4e) are only slightly affected by the rectangular hole defect, except the wavefield enhancement observed near to it, which is due to the stress concentration (42, 43). In addition, we have experimentally calculated the skyrmion number for the central skyrmion lattice in structures I and II, marked as cases $C_1$ and $C_2$, respectively. These experimental results, shown in Fig. 4c, indicate very small changes in the skyrmion number compared with that of the reference structure (case $C_3$), which confirms that the phononic skyrmions are robust against the defects.

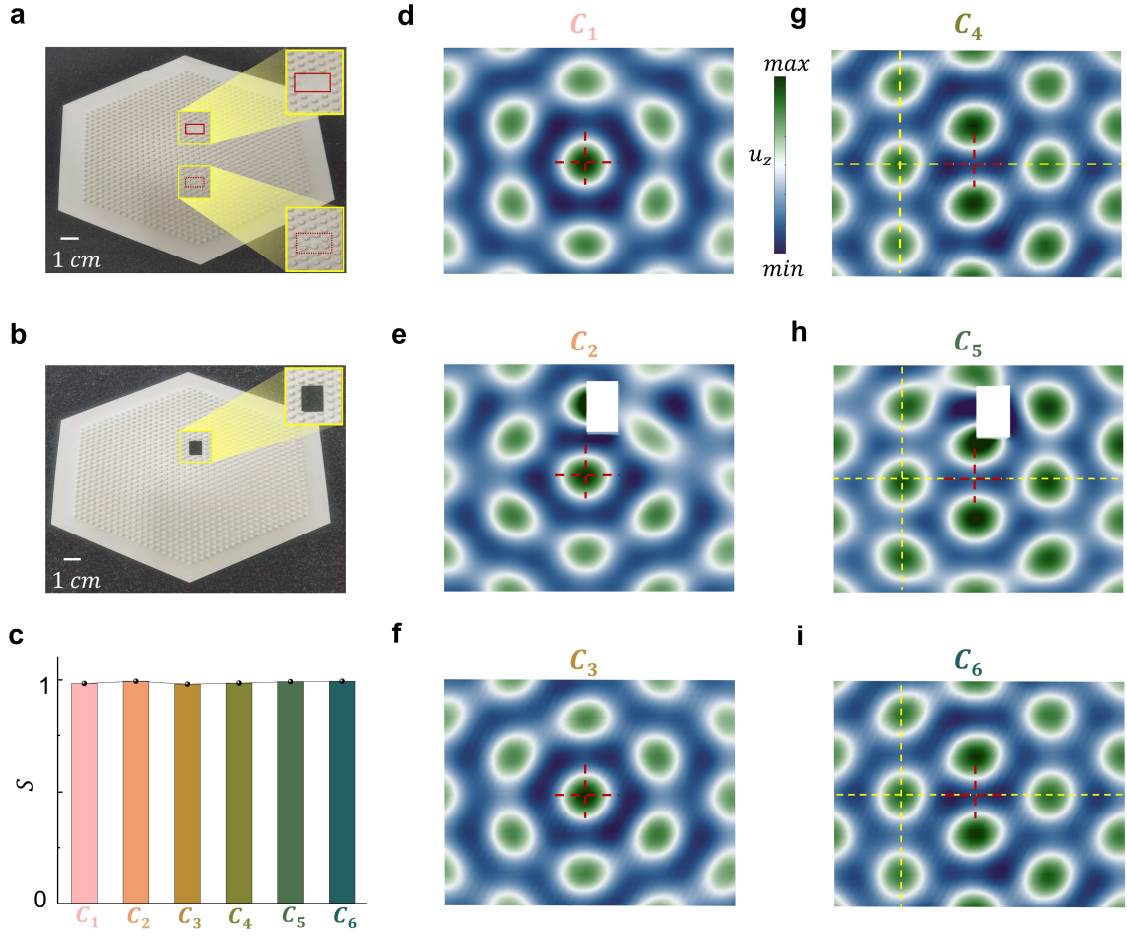

Fig. 4. **Robustness of phononic skyrmions. a**, The structure with the defects of removing (see the solid line box) and disorderly arranging (see the dashed line box) pillars, i.e., structure I. **b**, The structure with a rectangular hole, i.e., structure II. **c**, The experimental skyrmion numbers for the cases from $C_1$ to $C_6$. **d-f**, Experimental axial wavefields at the center of structure I, structure II, and reference structure without defects, respectively, marked as the cases from $C_1$ to $C_3$. **g-i**, Experimental wavefields in cases from $C_1$ to $C_3$ are shifted leftward by $\lambda$, respectively, marked as the cases from $C_4$ to $C_6$. The red dotted line intersection is the midpoint of the central lattice before it shifts. Distance between the yellow dotted line intersection and the red one is $\lambda$.

The phonic skyrmion lattices are created from the spin field interference of excited plane Lamb waves in the hexagonal-symmetry meta-plate. According to the coordinate transformation in the geometric space, the topologically protected wavefield can be flexibly moved by tuning the phase of the 3 pairs of excitation sources, even in the presence of defects. To move the skyrmion lattice by $\Delta \mathbf{r} = (\Delta x, \Delta y)$, we only need to tune the phase of three pairs of excitation sources by $\Delta \phi_{\theta = -\frac{\pi}{3}, 0, \frac{\pi}{3}} = k \cdot \Delta \mathbf{r} \cdot \mathbf{t}(\theta)$. For one of the simplest cases of shifting leftward $\lambda$, i.e., $\Delta \mathbf{r} = (\lambda, 0)$, the modulated phases are $\left( \Delta \phi_{\theta = 0} \quad \Delta \phi_{\theta = -\pi/3} \quad \Delta \phi_{\theta = \pi/3} \right) = \left( 0 \quad \pi \quad \pi \right)$. Therefore, we just need to tune the phase of excitation sources in the directions of $\theta = -\pi/3$ and $\theta = \pi/3$ by $\pi$. Figures 4g, 4h, and 4i provide the measured wavefields in Figs. 4d, 4e, and 4f shifted leftward by $\lambda$, marked as cases $C_4$, $C_5$, and $C_6$, respectively. Their experimentally calculated skyrmion numbers, shown in Fig. 4c, all approaches one. This confirms that the excitation source phase is an effective degree of freedom to manipulate the topological skyrmion configuration, even in the presence of defects.

**Conclusion and outlook**

We have theoretically and experimentally demonstrated and observed the formation of ultra-broadband phononic skyrmions based on the three-dimensional hybrid spin of elastic waves. The phononic skyrmion provides an unprecedented topological phononic structure, which has robustness against local defects of disorder, sharp corners, and even rectangular holes due to the nontrivial topological protection. The three-dimensional spin states open new pathways for topological phonons manipulation and may facilitate the exploration of new topological orders.

Our research opens possibilities to realize novel topological phononic materials in different scales from macro to microstructures. Although our reported results here concern frequencies around 8 kHz, the dimensionless dispersions shown in Fig. S2 allow our established skyrmion configuration to be readily scaled to higher frequencies (see constructed phononic skyrmion field on a silicon chip at a frequency of 1.5 MHz in Fig. S13). For example, reducing the plate thickness $h$ of 1 mm in our model to 100 nm (chip-scale level) and maintaining the same magnitude order of material parameters will raise operating frequencies to near 80 MHz, which could be used for high-frequency signal-processing chip-scale(17) technology with topological protection. The latter can overcome the difficulty of sensitivity of the conventional chip-scale technology to defects in high frequencies.

The phononic skyrmion can be created in any elastic wave system with a large impedance-mismatch solid-solid interface, solid-gas interface, or solid-liquid interface. The reason is that the longitudinal and transverse waves will hybrid at these interfaces to form elastic spin fields (31). The latter can be used to construct the phononic skyrmion configuration. Especially for the solid-liquid interface, phononic skyrmions constructed on the solid surface can induce skyrmions in fluids through fluid-solid coupling, which promotes the excitation and interaction of wavefield information in multiple fields. This fluid-solid coupling also paves the way toward skyrmion lattices "on-demand" for matter systems in fluids (e.g., cell manipulation in microfluidics (44), acoustofluidic pump (45), and biomedical testing (46)).

Finally, efficient transduction between elastic and electromagnetic waves can be achieved using piezoelectric coupling with microwaves (47-49) and optomechanical coupling with light (50). These couplings between different physical fields may give rise to a high-degree-of-freedom topological skyrmion system, and facilitate an advanced integrated information platform of interdisciplinary physics among electronics, photonics, and phononics.

**Method**

<u>Structure fabrication and materials</u>

Our established paradigm for generating and manipulating phononic skyrmions is universal and applicable to almost all solid materials. Here, we have chosen a 3D printer to fabricate our structures. 3D printing allows us to easily fabricate phonon structures with any geometry, providing an excellent platform for exploring and verifying the physics of phonons. The printed meta-plate is shown in Fig. 3a. The thickness of the hosting plate below pillar resonators is $h=1$ mm. The height and radius of pillar resonators are $3h$ and $h$, respectively. The lattice constant of the single-unit model is $a_0 = 4h$. The 3D printed material is PLA. Its material parameters were tested in our previous work(40). Young's modulus, Poisson's ratio, and the density are $E_{\mathrm{PLA}} = 3.44$ GPa , $\upsilon_{\mathrm{PLA}} = 0.35$ , and $\rho_{\mathrm{PLA}} = 1086.3$ kg/m$^3$ , respectively. The material of all the studied models in this work is PLA, except as noted in particular.

<u>Experiments</u>

As shown in Fig. 3a, we have attached six PZT patch arrays (a total of 36 PZT patches) to the structure surface in three directions of $\theta = 0°, \ 60°$, and $-60°$. These PZT patch arrays are driven by a signal generator (Tektronix AFG3022C) to excite plane lamb waves. The surface of the structure back (the surface without pillared resonators) is held perpendicular to the laser beam from the PSV-500 laser vibrometer. The PSV-500 scanning head records the axial displacements of all measurement points in the far-field. An ensemble average with 20 samples is used at every measurement point to ensure signal quality. For the measured 3D time-resolved vector field shown in Moves S1-S3, the excitation signal is a 50-cycle tone burst $w_t = A_0 \left[ 1 - \cos\left(2\pi f_c t \, / \, 50\right)\right]\sin\left(2\pi f_c t\right)$, where $f_c = 8$ kHz is the central frequency. The sampling frequency in the time domain was set at 125 kHz, and measure resolution is 8 us. The images are processed to create vector representations of the wavefield, which in turn are combined in a time sequence to create a video of the time development of the wavefield vectors.

<u>Numerical simulations</u>

The full-wave simulations were implemented by the commercial software COMSOL Multiphysics with the Solid Mechanics Module and Acoustic-Solid Interaction module, based on the finite element method. An eigenfrequency study was used to simulate the band-like dispersion. The largest mesh element size was set lower than one-twelfth of the lowest wavelength, and finer meshes were applied at the region with pillars. The parameters in the simulations are consistent with those of the experimental structures.

**Acknowledgments:**

The authors acknowledge Professor Ming-Hui Lu for fruitful discussions.

**Funding:**

Air Force Office of Scientific Research under award number FA9550-18-1-7021

la Région Grand Est and Institut Carnot ICEEL

**Author contributions:**

B. A. and L. C. proposed the concept. L. C. and B. A. proposed the methodology and approaches. L. C. and S. W. conceived and designed the experiments. L. C. and B. A. wrote the paper. B.A. guided the research. All the authors contributed to data analysis and discussions.

**Competing interests:**

Authors declare that they have no competing interests.

**Data and materials availability:**

All data are available in the main text or the supplementary materials.

**Supplementary Video 1 The 3D time-resolved spin vector field**

We only turn on one piezoelectric patch array (i.e., $1^{\#}$ array in Fig. 3a) to excite a plane A0 wave propagating along the $x$-axis directions. The dynamic field in black and white (black, positive; white, negative) beneath the vectors is created from the original data of the measured displacements by PSV-500. The data covers the delay time from $\tau = 22.064$ ms to 22.480 ms.

**Supplementary Video 2 The 2D time-resolved spin vector field**

The 2D time-resolved spin vector field is a subset of the vector data of Video 1 in the *x-z* plane.

**Supplementary Video 3 The 3D time-resolved Néel-type skyrmion vector field**

The central skyrmion lattice consists of 726 measured points, with 16 images over a wave cycle, each of size 2156 by 1580 pixels. A full-wave cycle takes 0.125 ms, and the measurement resolution is 8 us. The data covers the delay time from $\tau = 22.064$ ms to 22.480 ms. At 6.67 frames per second, a skyrmion wavefield period in the video oscillates every 2.4 seconds.